\documentclass[aps,twocolumn,showpacs]{revtex4}
\usepackage{amssymb}
\usepackage{graphicx}

\renewcommand{\ss}{\scriptscriptstyle}
\begin{document}

\title{Modulation of intersubband IR absorption under intense THz irradiation%
}
\author{A. Hern{\'{a}}ndez-Cabrera}
\email{ajhernan@ull.es}
\author{P. Aceituno}
\affiliation{Dpto. F{\'\i}sica B\'{a}sica, Universidad de La Laguna \\
La Laguna, 38206-Tenerife, Spain}
\author{F.T. Vasko}
\affiliation{Institute of Semiconductor Physics, NAS Ukraine \\
Prospekt Nauki 41, Kiev, 03028, Ukraine}
\date{\today}

\begin{abstract}
We analyze the modification of the intersubband absorption of electrons in
quantum wells under intense THz irradiation. An expression for the induced
current is obtained, based on the adiabatic approach and the resonant
approximation. We predict the occurrence of a significant fine structure as
well as the broadening and shift of the absorption under THz pump in a MW/cm$%
^2$ intensity range.
\end{abstract}

\maketitle

%\email{ftvasko@yahoo.com}

\section{Introduction}

Strong transverse fields have long been known for modifying confined states
in quantum wells (QWs). The examination of the interband optical transitions
under transverse fields, both static and high-frequency, is a convenient
method to study these modifications (see references in \cite{1,2} and \cite%
{3,4}, respectively). The excitonic effect and the modifications of both
electron and hole states under transverse fields have to be taken into
account for a quantitative description of the interband linear response. It
is also interesting to study the intersubband response under infrared (IR)
excitation of electrons between the ground and the excited conduction band
states which are placed in a transverse field. Whereas the electro-optic
modulation of the intersubband transitions is well investigated \cite{5},
the influence of an intense THz irradiation on such transitions is not
investigated to the best of our knowledge. In this paper we treat
theoretically the effect of the THz pump on the IR intersubband absorption.

The confined electron states in a QW, of width $d$, subjected to a
transverse electric field $E_{\omega }\cos \omega t$, are described within
the adiabatic approach, if $\hbar \omega \ll \varepsilon _{21}$, where $%
\varepsilon _{21}/\hbar $ is the frequency of the intersubband transitions.
Since the levels oscillate with a frequency $\omega $, the $(n+1)$-order
intersubband transitions, with $n$ THz photons and a single IR photon, take
place resulting in a fine structure of the absorption. At the same time, the
shape of the absorption peaks is modified under the THz irradiation. In
addition, one may consider the THz irradiation as a perturbation if $%
|e|E_{\omega }d/2<\varepsilon _{21}$. Otherwise a numerical description of
the electron states have to be applied.

\begin{figure}[tbp]
\begin{center}
\includegraphics{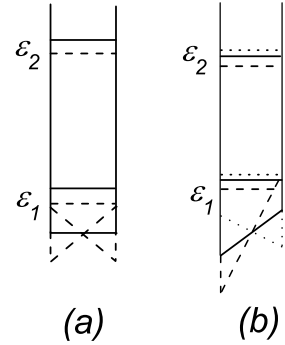}
\end{center}
\par
\addvspace{-0.5 cm}
\caption{Band diagramms for symmetric $(a)$ and non-symmetric $(b)$ QWs
under transverse THz irradiation. Dashed and dotted lines schematically show
variation of levels and potentials under maximal THz field.}
\label{fig.1}
\end{figure}

The calculations below are based on the one-particle density matrix equation
linearized with respect to the IR field $E_{\Omega }\exp (-i\Omega t)$,
while the THz irradiation is taken into account in the framework of the
adiabatic approach. The broadening is described by the phenomenological
approach which takes into account the $LO$-phonon emission in the spectral
region $\Omega >\omega _{LO}$, where $\omega _{LO}$ is the optical phonon
frequency. We have considered two cases: a symmetric rectangular QW, and a
non-symmetric one. The last case may be realized by adding a transversal dc\
electric field $E_{o}$ to the symmetric case, as can be seen in Fig. 1$(a)$
and 1$(b)$, respectively.

The paper is organized as follows. In Sec. II we derive the relative
intersubband absorption under THz pump starting on the density matrix
equation. The case of the perturbative approach is considered in Sec. III,
while the results for the numerical description are discussed in Sec. IV.
Concluding remarks and a list of the assumptions made are given in the last
section.

\section{Intersubband response}

In this section, we obtain the averaged over the THz pump period relative
absorption of the IR probe. The high-frequency addendum to the density
matrix $\Delta \rho _{t}(z,z^{\prime })\exp (-i\Omega t)$ is governed by the
linearized equation written in the coordinate-momentum representation: 
\begin{eqnarray}
\frac{\partial \Delta \rho _{\mathbf{p}t}(z,z^{\prime })}{\partial t}+%
\frac{i}{\hbar }(\hat{h}_{zt}-\hat{h}_{z^{\prime }t}-\hbar \Omega )\Delta
\rho _{\mathbf{p}t}(z,z^{\prime })  \nonumber \\
+\frac{i}{\hbar }(\widehat{\delta h}_{z}-\widehat{\delta h}_{z^{\prime
}})\rho _{\mathbf{p}t}(z,z^{\prime })=0
\end{eqnarray}%
with the transverse coordinate $z$ and 2D momentum $\mathbf{p}$. Here the
Hamiltonian and the perturbation operator, $\hat{h}_{zt}$ and $\widehat{%
\delta h}_{z}\exp (-i\Omega t)$, are given by 
\begin{equation}
\hat{h}_{zt}=\frac{\hat{p}_{z}^{2}}{2m}+w_{z}+eE_{\omega }z\cos \omega t,~~~~%
\widehat{\delta h}_{z}=\frac{ie}{\Omega }E_{\Omega }\hat{v}_{z},
\end{equation}%
where $\hat{v}_{z}=\hat{p}_{z}/m$ is the velocity operator and $w_{z}$ is
the potential energy inside the QW. Within the hard-wall approximation one
have to consider the interval $|z|<d/2$ with the zero boundary conditions.
The density matrix under THz irradiation satisfies the periodicity
condition: $\rho _{\mathbf{p}t+2\pi /\omega }(z,z^{\prime })=\rho _{\mathbf{p%
}t}(z,z^{\prime })$.

It is convenient to use the parametrically time-dependent wave functions, $%
\varphi _{zt}^{k}$, determined by the eigenstate problem: 
\begin{equation}
\hat{h}_{zt}\varphi _{zt}^{k}=\varepsilon _{kt}\varphi _{zt}^{k},~~~~\varphi
_{z=\pm d/2,t}^{k}=0,
\end{equation}%
which is defined in the interval $|z|<d/2$. The expansion of the first and
zero-order density matrices $\Delta \rho _{\mathbf{p}t}(z,z^{\prime })$ and $%
\rho _{\mathbf{p}t}(z,z^{\prime })$ over this basis gives us: 
\begin{eqnarray}
\Delta \rho _{\mathbf{p}t}(z,z^{\prime }) &=&\sum_{kk^{\prime }}\Delta \rho
_{\mathbf{p}t}(k,k^{\prime })\varphi _{zt}^{k\ast }\varphi _{zt}^{k^{\prime
}},  \nonumber \\
\rho _{\mathbf{p}t}(z,z^{\prime }) &=&\sum_{kk^{\prime }}\rho _{\mathbf{p}%
t}(k,k^{\prime })\varphi _{zt}^{k\ast }\varphi _{zt}^{k^{\prime }}
\end{eqnarray}%
and Eq. (1) in the $k$-representation takes the form: 
\begin{eqnarray}
&&\frac{\partial \Delta \rho _{\mathbf{p}t}(k,k^{\prime })}{\partial t}+%
\frac{i}{\hbar }(\varepsilon _{kt}-\varepsilon _{k^{\prime }t}-\hbar \Omega
)\Delta \rho _{\mathbf{p}t}(k,k^{\prime })  \nonumber \\
&&+\sum_{k^{\prime }}\left[ \Delta \rho _{\mathbf{p}t}(k,k^{\prime })\delta
_{t}^{k^{\prime }k}-\delta _{t}^{kk^{\prime }}\Delta \rho _{\mathbf{p}%
t}(k^{\prime },k)\right]  \nonumber \\
&=&\frac{i}{\hbar }\sum_{k^{\prime }}\left[ \delta h_{t}^{kk^{\prime }}\rho
_{\mathbf{p}t}(k^{\prime },k^{\prime })-\rho _{\mathbf{p}t}(k,k^{\prime
})\delta h_{t}^{k^{\prime }k^{\prime }}\right] ,
\end{eqnarray}%
where the non-adiabatic factor $\delta _{t}^{kk^{\prime }}$ and the
perturbation $\delta h_{t}^{kk^{\prime }}$ are given by: 
\begin{eqnarray}
\delta _{t}^{kk^{\prime }} &=&\int_{-d/2}^{d/2}dz\frac{\partial \varphi
_{zt}^{k\ast }}{\partial t}\varphi _{zt}^{k^{\prime }},  \nonumber \\
\delta h_{t}^{kk^{\prime }} &=&\frac{ie}{\Omega }E_{\Omega
}\int_{-d/2}^{d/2}dz\varphi _{zt}^{k\ast }\hat{v}_{z}\varphi
_{zt}^{k^{\prime }}.
\end{eqnarray}%
We have taken into account that $\delta _{t}^{kk^{\prime }}=-\delta
_{t}^{k^{\prime }k~\ast }$.

The induced current density is written through $\Delta \rho $ as follows: 
\begin{eqnarray}
j_{t} &=&\frac{e}{L^{2}}\sum_{\mathbf{p}}\int_{-d/2}^{d/2}dz%
\lim_{z_{1},z_{1}^{\prime }\rightarrow z}(\hat{v}_{z_{1}}+\hat{v}%
_{z_{1}^{\prime }})\Delta \rho _{\mathbf{p}t}(z_{1},z_{1}^{\prime }) 
\nonumber \\
&=&\frac{2e}{L^{2}}\sum_{\mathbf{p}kk^{\prime }}v_{k^{\prime }k}(t)\Delta
\rho _{\mathbf{p}t}(k,k^{\prime }),
\end{eqnarray}%
where $L^{2}$ is the normalization area and $v_{kk^{\prime
}}(t)=\int_{-d/2}^{d/2}dz\varphi _{zt}^{k\ast }\hat{v}_{z}\varphi
_{zt}^{k^{\prime }}$. For the case of resonant excitation between the first
and the second levels, when $\hbar \Omega \sim \varepsilon _{21}$, Eq. (7)
can be transformed into $j_{t}\simeq ev_{12}(t)\Delta \rho _{t}(2,1)$. Here
we have also performed the summation over 2D momentum according to $\Delta
\rho _{t}(2,1)=(2/L^{2})\sum_{\mathbf{p}}\Delta \rho _{\mathbf{p}t}(2,1)$.
Using Eq. (5) and supposing that only the ground level is populated, we
obtain an equation for $\Delta \rho _{t}(2,1)$ in the form: 
\begin{eqnarray}
&&\frac{\partial \Delta \rho _{t}(2,1)}{\partial t}+i\left[ \Omega
_{21}(t)-\Omega -i\nu _{21}(t)\right] \Delta \rho _{t}(2,1)  \nonumber \\
&=&\frac{eE_{\ss \Omega }}{\hbar \Omega }v_{21}(t)n_{2D},
\end{eqnarray}%
where $\Omega _{21}(t)=(\varepsilon _{2t}-\varepsilon _{1t})/\hbar $ is the
time-dependent interlevel frequency, $\nu _{21}(t)=\nu _{\Omega }-\delta
_{t}^{22}+\delta _{t}^{11}$ is the effective relaxation frequency with the
phenomenological relaxation frequency $\nu _{\Omega }$, and $n_{2D}$ is the
electron concentration. The stepped frequency $\nu _{\Omega }$ increases in
the active region, $\Omega >\omega _{LO}$, due to the emission of LO-phonons
with the frequency $\omega _{LO}.$

Introducing the time-dependent conductivity tensor $\sigma _{{\Omega }%
t}=j_{t}/E_{\Omega }$ and solving Eq. (8), we write: 
\begin{eqnarray}
\sigma _{{\Omega }t} &=&\frac{e^{2}n_{2D}}{\hbar \Omega }v_{21}(t)\int_{-%
\infty }^{t}dt^{\prime }v_{21}(t^{\prime }) \\
&&\times \exp \left\{ -i\int_{t^{\prime }}^{t}d\tau \left[ \Omega _{21}(\tau
)-\Omega -i\nu _{21}(\tau )\right] \right\} .  \nonumber
\end{eqnarray}%
The averaged over the THz period relative absorption is determined as
follows \cite{6}: 
\begin{equation}
\xi _{\Omega }=\frac{4\pi }{c\sqrt{\epsilon }}Re\int_{-\pi /\omega }^{\pi
/\omega }\frac{dt}{2\pi /\omega }\sigma _{\Omega t}.
\end{equation}%
Thus, in order to calculate $\xi _{\Omega }$ one needs to find $\Omega
_{21}(t)$ and $\nu _{21}(t)$ and to perform the integrations over time.

\section{Parabolic Approach}

The perturbative solution of the eigenstate problem (3) can be written as
follows: 
\begin{eqnarray}
\varphi _{zt}^{k} &\simeq &\varphi _{z}^{k}+\sum_{k^{\prime }\neq k}\frac{%
eE_{\omega }z_{kk^{\prime }}}{\varepsilon _{k}-\varepsilon _{k^{\prime }}}%
\varphi _{z}^{k^{\prime }}\cos \omega t , ~~~ \\
\varepsilon _{kt} &\simeq &\varepsilon _{k}+eE_{\omega }z_{kk}\cos \omega
t+\sum_{k^{\prime }\neq k}\frac{|eE_{\omega }z_{kk^{\prime }}|^{2}}{%
\varepsilon _{k}-\varepsilon _{k^{\prime }}}\cos ^{2}\omega t,  \nonumber
\end{eqnarray}%
where the matrix elements $z_{kk^{\prime }}$ and the energy level $%
\varepsilon _{k}$ are determined from the zero-field eigenstate problem: $(%
\hat{p}_{z}^{2}/2m+w_{z})\varphi _{z}^{k}=\varepsilon _{k}\varphi _{z}^{k}$,
with the boundary conditions $\varphi _{|z|=d/2}=0$. The frequency $\Omega
_{21}(t)$ in Eq. (9) is written as follows: 
\begin{equation}
\Omega _ {21}(t)=\Omega _{21}+\Omega _{21}^{(1)}\cos \omega t+\Omega
_{21}^{(2)}\cos 2\omega t,
\end{equation}%
where the coefficients $\Omega _{21}^{(1,2)}$ are obtained from $\varepsilon
_{kt}$ given by Eq. (11). Note also, that $z_{kk}=0$ for the symmetric QW
and the time-dependent contribution to $\Omega _{21}(t)$ is $\propto \cos
2\omega t$ for such a case.

\begin{figure}[tbp]
\begin{center}
\includegraphics{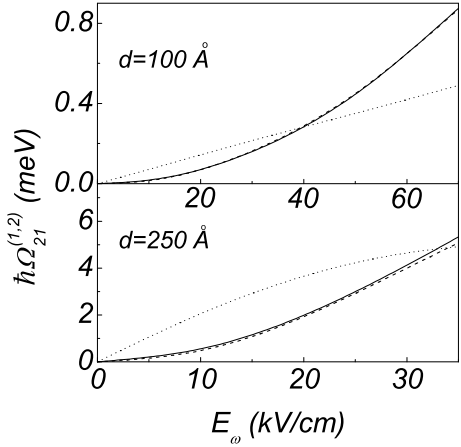}
\end{center}
\par
\addvspace{-0.5 cm}
\caption{The coefficients $\hbar \Omega _{21}^{(1,2)}$ versus THz field
strenth for narrow and wide QWs. Solid line: $\hbar \Omega _{21}^{(2)}$ for
symmetric case ($E_{0}=0$ kV/cm). Dashed (dotted) line: $\hbar \Omega
_{21}^{(2)}$ ($\hbar \Omega _{21}^{(1)}$) for non-symmetric case (QWs under
dc field $E_{o}=10$ kV/cm). }
\end{figure}

The parabolic and linear dependencies of $\Omega _{21}^{(2)}$ and $\Omega
_{21}^{(1)}$ on the applied THz field for both symmetric and asymmetric
cases are plotted in Fig. 2. The calculations are performed for narrow ($%
d=100$ $\mathring{A}$) and wide ($d=250$ $\mathring{A}$) $GaAs$-based QWs.
In order to model non-symmetric QW we have introduced a transverse dc field $%
E_{o}=10\ kV/cm$ besides the THz field $E_{\omega }\cos \omega t$. We can
see in Fig. 2 that the behavior of $\Omega _{21}^{(2)}$ coefficients is
exactly the same for the symmetric and non-symmetric cases in the region
where the parabolic approximation is valid ( $|e|E_{\omega }d/2<\varepsilon
_{21}$). That is, for the narrow QW a parabolic bearing can be found up to
very strong fields ($E_{\omega }\sim 300$ kV/cm); whereas, for the wide QW,
this approximation is only valid up to $E_{\omega }\sim 20$ kV/cm and beyond
these regions $\Omega _{21}(t)$ cannot be written by means of Eq. (12). For
the non-symmetric case, $\Omega _{21}^{(1)}$ coefficients show an almost
linearly increase in the parabolic approximation range for both narrow and
wide QWs. After that, $\Omega _{21}^{(1)}$ tend to a constant value.

The non-adiabatic factors appearing in $\nu _{21}(t)$ are given by 
\begin{equation}
\delta _{t}^{kk}\simeq -\frac{\omega }{2}\sum_{k^{\prime }\neq k}\frac{%
|eE_{\omega }z_{kk^{\prime }}|^{2}}{(\varepsilon _{k}-\varepsilon
_{k^{\prime }})^{2}}\sin 2\omega t\equiv \delta _{kk}\sin 2\omega t.
\end{equation}%
Thus, $\nu _{21}(t)$ can be written through $\Delta \nu \equiv \delta
_{22}-\delta _{11}$. The velocity matrix element modifies weakly with time, $%
v_{21}(t)\simeq v_{21}$, because the time-dependent contributions ($\propto
\cos \omega t,~\cos 2\omega t$) are two order smaller than $v_{21}$. For the
structures under consideration $v_{21}\simeq 4\times 10^{7}\ cm/s$ ($d=100$ 
\AA ) and $v_{21}\simeq 2\times 10^{7}\ cm/s$ ($d=250$ \AA ). The
non-adiabatic factor $\Delta \nu $ also shows a parabolic behavior which
slightly depends on $E_{o}$. For the region of parameters considered here, $%
\Delta \nu /\omega $ varies from $0$ to $5\times 10^{-3}$ and we have
neglected these contributions below.

Using the expansion of the exponential factors in (9) over the Bessel
functions \cite{6} and performing the integrations over time, we obtain $\xi
_{\Omega }$ for the symmetric QW in the form: 
\begin{equation}
\xi _{\Omega }\simeq \frac{4\pi }{\sqrt{\epsilon }}\frac{e^{2}}{\hbar c}%
\frac{\nu |v_{21}|^{2}}{\Omega }n_{2D}\sum_{k=-\infty }^{\infty }\frac{%
J_{k}(\Omega _{21}^{(2)}/4\omega )^{2}}{(\Omega _{21}-2k\omega -\Omega
)^{2}+\nu ^{2}}.
\end{equation}%
In Fig. 3 we have plotted the spectral dependencies of $\xi _{\Omega }$,
given by Eq. (14), under different THz pumps for the narrow symmetric QW.
The calculations are performed for a multiple quantum well structure with
ten decoupled 100 \AA\ wide $GaAs$ QWs and for THz quanta energy values $%
\hbar \omega =0.5$ meV and $1.5$ meV. We have used two broadening energy
values $\hbar \nu \simeq 0.66$ meV (corresponding to a relaxation frequency $%
\nu =1$ ps$^{-1}$) and $\hbar \nu =1.5$ meV. One can see that, for fields
well below the parabolic approximation limit, the relative absorption shows
a monotonous decreasing while spreads to higher frequencies. For high
fields, beyond $70$ kV/cm, a clear multi-peak structure appears because the
strong-modulation case, when $\Omega _{21}^{(2)}/4\omega >1$, can be
realized in the framework of the perturbation approach. The relative height
of the different peaks for this structure depends on the quanta energy
value, as can be seen from Fig. 3, due to the $\omega $\ contribution to the
Bessel functions and the denominators in Eq. (14). Panels (a) and (c), with $%
0.5$ meV, show higher lateral peaks than the central one, whereas panels (b)
and (d), with $1.5$ meV, have a clear maximum in a central peak. The
relaxation energy not only affect the broadening of the peaks but also the
relative absorption and so, panels (c) and (d) with $1.5$ meV show wider and
lower peaks than panels (a) and (b) in which the fine structure is more
evident and the relative absorption is about twice bigger than that of the
other cases. 
\begin{figure}[tbp]
\begin{center}
\includegraphics{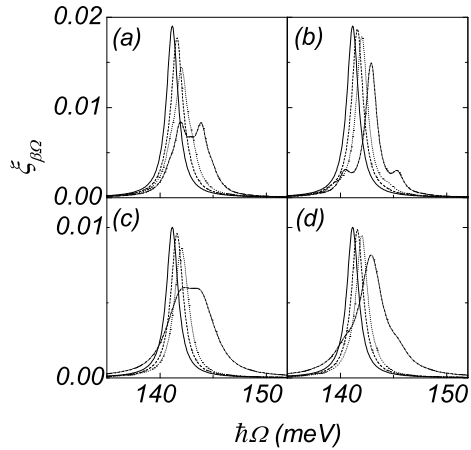}
\end{center}
\par
\addvspace{-0.5 cm}
\caption{Spectral dependencies of $\protect\xi _{\Omega }$ versus $\hbar
\Omega $ for the symmetric case under THz field strengths: $E_{\protect%
\omega }=0$ kV/cm (solid line), $E_{\protect\omega }=50$ kV/cm (dashed
line), $E_{\protect\omega }=70$ kV/cm (dotted line) and $E_{\protect\omega %
}=100$ kV/cm (dot-dashed line). Panel (a): $\hbar \protect\omega =0.5$ meV, $%
\hbar \protect\nu =0.66$ meV; panel (b): $\hbar \protect\omega =1.5$ meV, $%
\hbar \protect\nu =0.66$ meV; panel (c): $\hbar \protect\omega =0.5$ meV, $%
\hbar \protect\nu =1.5$ meV; panel (d): $\hbar \protect\omega =1.5$ meV, $%
\hbar \protect\nu =1.5$ meV.}
\label{fig.3}
\end{figure}

There are more sums over $k$ for the non-symmetric QW due to the additional
term with coefficient $\Omega _{21}^{(1)}$. In this case the parabolic
approximation leads to 
\begin{eqnarray}
\xi _{\Omega } \simeq \frac{4\pi }{\sqrt{\epsilon }}\frac{e^{2}}{\hbar c}%
\frac{\nu |v_{21}|^{2}}{\Omega }n_{2D}\sum_{kk_{1}\Delta k=-\infty }^{\infty
}~~~~~ \\
\times \frac{J_{k_{1}}\left( \frac{\Omega _{21}^{(1)}}{\omega }\right)
J_{k_{1}-2\Delta k}\left( \frac{\Omega _{21}^{(1)}}{\omega }\right)
J_{k+\Delta k}\left( \frac{\Omega _{21}^{(2)}}{4\omega }\right) J_{k}\left( 
\frac{\Omega _{21}^{(2)}}{4\omega }\right) }{[\Omega _{21}-(2k-k_{1})\omega
-\Omega ]^{2}+\nu ^{2}}.  \nonumber
\end{eqnarray}%
Fig. 4 shows the relative absorption of non-symmetric QW for $E_{o}=10$
kV/cm and different $E_{\omega }$ values. In order to obtain a higher
relative absorption and a clearer peak structure, we have used parameters
corresponding to panel (b) of Fig. 3 ($\hbar \omega =1.5$ meV and $\hbar \nu
=0.66$ meV). The behavior of the relative absorption is similar to that of
the corresponding to the symmetric case for low fields, but the relative
absorption decreases faster and, for fields beyond $70$ kV/cm, when the
multi-peak fine structure becomes evident, there are clear differences
between this case and the symmetric one. Looking at the $100$ kV/cm case in
panel (b) of Fig. 3, we can see the symmetry of the relative absorption
multi-peak structure with a central maximum and two lateral satellites. On
the other hand, Fig. 4 shows a clear non-symmetric fine structure, with a
pronounced central maximum located at the same energy than the corresponding
to the symmetric case. However, lateral satellites show a strong asymmetry.
At first glance it seems that a new plateau \ appears in the low energy
side. Actually, for $E_{\omega }=$ $100$ kV/cm, both symmetric and
non-symmetric cases present seven relative maxima as can be proved by
diminishing the relaxation broadening beyond realistic values. In other
words, the absorption multi-peak structure is masked in part by both the THz
quanta and the effective relaxation energy values.

\begin{figure}[tbp]
\begin{center}
\includegraphics{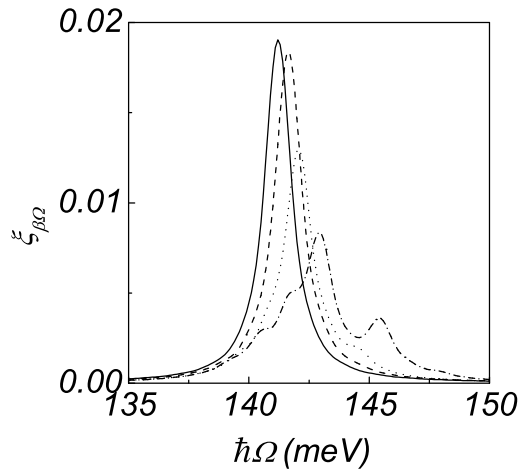}
\end{center}
\par
\addvspace{-0.5 cm}
\caption{Relative absorption for non-symmetric $100$ \AA\ QW case under $%
E_{o}=10$ kV/cm and $E_{\protect\omega }=0$ (solid line), $E_{\protect\omega %
}=50$ kV/cm (dashed line), $E_{\protect\omega }=70$ (dotted line) and $E_{%
\protect\omega }=100$ kV/cm (dot-dashed line). $\hbar \protect\omega =1.5$
meV and $\hbar \protect\nu =0.66$ meV.}
\label{fig.4}
\end{figure}

\section{Numerical description}

As we have already noted, the perturbation approach is valid under the
condition $|e|E_{\omega }d/2<\varepsilon _{21}$. If the THz pump is
stronger, one needs to perform a numerical solution for the eigenstate
problem (3) and to integrate Eqs. (9, 10) numerically taking into account
the time-dependent velocity matrix element. Such a case can be realized in a
wide QW without the requirement of using high fields. Therefore, we will
consider below $250$ \AA\ wide $GaAs$-based QWs. In order to check the
validity of the parabolic approximation, Fig. 5 shows the relative
absorption calculated both by means of this approach and numerically.
Calculations are performed for a multiple QW structure with ten decoupled
QWs and for a\ THz quanta energy $\hbar \omega =1.5$ meV. Due to the low
values of the energy range in this case (around the optical phonon energy),
we consider a step-like function for the relaxation energy taking a smaller
value $\hbar \nu =0.5$ meV in the passive region, when $\hbar \Omega <35$
meV, and a bigger one $\hbar \nu =1.5$ meV for higher energy values. One can
see a very good correspondence between the two methods for fields up to $30$
kV/cm while, for higher fields, the parabolic approach clearly fails.

\begin{figure}[tbp]
\begin{center}
\includegraphics{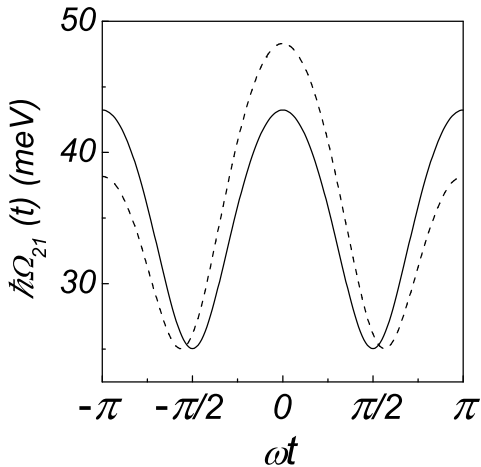}
\end{center}
\par
\addvspace{-0.5 cm}
\caption{Relative absorption $\protect\xi _{\Omega }$ obtained through
numerical calculations (solid line: $E_{\protect\omega }=30$ kV/cm, dotted
line: $E_{\protect\omega }=50$ kV/cm) and by means of the parabolic
approximation (dashed line: $E_{\protect\omega }=30$ kV/cm, dash-dotted
line: $E_{\protect\omega }=50$ kV/cm) for the symmetric case and wide QW.}
\end{figure}

The results obtained by means of the numerical description for spectral
dependencies beyond $E_{\omega }=30$ kV/cm are represented in Figs. 6
(symmetric QW) and 7 (non-symmetric QW). As in Fig. 5, calculations have
been made for ten QWs and for the same THz quanta energy and step-like
relaxation energy. For higher THz fields the asymmetry appears (even for the
symmetric QW). Together with the decreasing and spreading of the relative
absorption, the initial peak for $E_{\omega }=0$ kV/cm gradually splits in
several peaks as the field intensity increases, the number of peaks
depending on the intensity. The non-symmetric QW well case shows two
different regions not only due to the stepped relaxation but also because
the field runs twice per cycle the region between $0$ and $E_{o}-E_{\omega }$
and only once the remaining region. Thus, the multi-peak fine structure is
partially hidden for $\hbar \Omega >35$ meV due, once again, to the higher
effective relaxation energy. 
\begin{figure}[tbp]
\begin{center}
\includegraphics{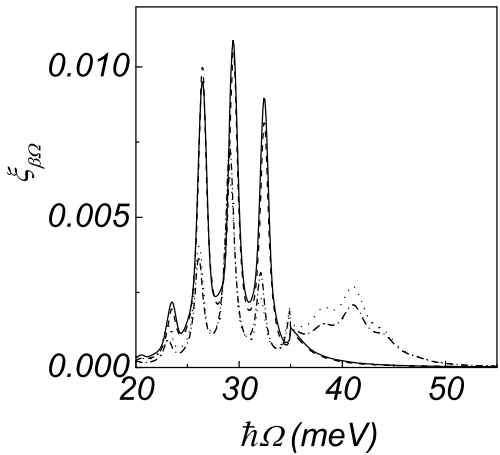}
\end{center}
\par
\addvspace{-0.5 cm}
\caption{Symmetric wide QW spectral dependencies of $\protect\xi _{\Omega }$
versus $\hbar \Omega $ for $E_{\protect\omega }=30$ kV/cm (solid line), $50$
kV/cm (dashed line), and \ $70$ kV/cm (dotted line). }
\end{figure}

\bigskip 
\begin{figure}[tbp]
\begin{center}
\includegraphics{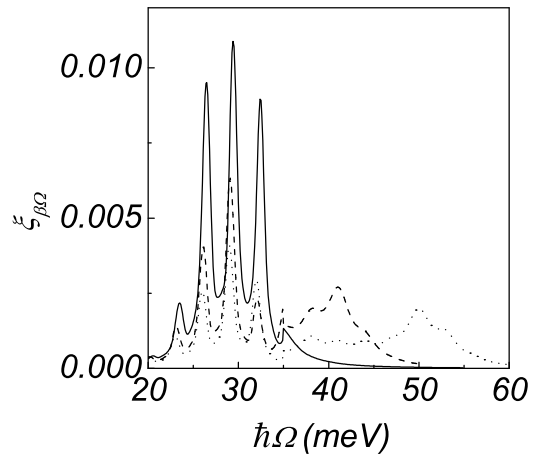}
\end{center}
\par
\addvspace{-0.5 cm}
\caption{Non-symmetric QW spectral dependencies of $\protect\xi _{\Omega }$
versus $\hbar \Omega $ for $E_{0}=10$ kV/cm and $E_{\protect\omega }=30$
kV/cm (solid line), $50$ kV/cm (dashed line), $70$ kV/cm (dotted line).}
\label{fig.7}
\end{figure}

\section{Conclusions}

We have studied the modifications of the intersubband absorption in a
multi-QW structure caused by a strong THz irradiation. Experimentally, such
an irradiation can be achieved by using free-electron or gas lasers with an
energy density in the MW/cm$^{2}$ range (see applications of these lasers to
study a heterostructure response in \cite{8} or \cite{9}, respectively).
Results show a significant fine structure of the absorption peak due to the $%
(n+1)$-order intersubband transitions, with $n$ THz photons and a single IR
photon. In addition, a strong modification of absorption, which consists on
a noticeable broadening of the zero-field peak and a shift towards higher
energy values, is also demonstrated. Since the relative absorption
calculated is weak enough, one can measure photoconductivity.

Next, we discuss the assumptions used in the above calculations. Within the
hard-wall scheme we have disregarded the underbarrier penetration. In
general, for deep and wide decoupled (or weakly coupled) quantum wells,
underbarrier penetration slightly modifies wave functions and energy levels
position. In the present case and due to the relatively big interlevel
distance ($\varepsilon _{21}\sim 140$ meV for narrow QW and $\varepsilon
_{21}\sim 30$ meV for wide QW) little changes in the level positions do not
affect essentially results. For the same reasons we have not taken into
account possible contributions of the Coulomb renormalization neglecting the
depolarization and exchange effects. We have also used a phenomenological
homogeneous broadening energy $\hbar \nu \sim 0.5-1.5$ meV. These
assumptions are generally accepted and a possible improvement will not
change essentially the obtained results. We have also neglected the
interlevel redistribution under THz pump because the interlevel energy $%
\varepsilon _{21}$ is bigger than $20$ meV while the THz quanta energy $%
\hbar \omega $ is around $1$ meV. One can neglect a THz field effect on the
damping because the influence of such a field on the wave functions is not
very strong. Thus, a phenomenological broadening should not be essentially
dependent on the THz pump.

To conclude, the present work has shown the possibility that an essential
modification of the IR intersubband response takes place when an intense THz
irradiation is applied. We expect that the results obtained encourages
researches to carry out experiments in this direction.


\begin{thebibliography}{99}
\bibitem{7} The Bessel-based expansions of the exponential factors used in
Sec. III are:

$\exp (iz\cos\psi)=\sum_{l=-\infty}^{\infty}i^le^{il\psi}J_l(z)$

$\exp (iz\sin\psi)=\sum_{l=-\infty}^{\infty}e^{il\psi}J_l(z)$

$\exp (z\cos\psi)=\sum_{l=-\infty}^{\infty}e^{il\psi}I_l(z)$
\end{thebibliography}
\end{document}